\documentclass[preprint,authoryear,12pt]{elsarticle}

\usepackage{epsfig}

\usepackage{amssymb}

\usepackage[ps2pdf,%
a4paper=true,%
breaklinks=true,%
colorlinks=true,%
pdfauthor={First Author et al.},%
pdftitle={Template for manuscripts in Advances in Space Research}%
]{hyperref}

\journal{Advances in Space Research}

\usepackage[nolist]{acronym}

\begin{document}

%%%%%%%%%%%%%%%%%%%%%%%%%%%%%%%%%%%%%%%%%%%%%%%%%%%%%%%%%%%%%%%%%%%%%%%%%%%%%
%% Frontmatter
\begin{frontmatter}


% Use the tnoteref command within \title and fnref within \author or \address for footnotes;
% use the corref command within \author for corresponding author footnotes;
% \author{Name\corref{cor1}\fnref{label2}}
% \address{Address\fnref{label3}}

\title{Torsional Balloon Flight Line Oscillations: Comparison of Modelling to Flight Data}%\tnoteref{footnote1}}
% \tnotetext[footnote1]{This template can be used for all publications in Advances in Space Research.}

% Use optional labels to link authors explicitly to addresses:

\author{Fran\c{c}ois Aubin\corref{cor}}%\fnref{footnote2}}
\address{University of Minnesota School of Physics and Astronomy, Minneapolis, MN 55455}
\cortext[cor]{Corresponding author}
% \fntext[footnote2]{Additional information regarding the corresponding author}
\ead{faubin@umn.edu}

% Url can be given like this:
% \ead[url]{http://www.elsevier.com/wps/find/authorsview.authors/latex}

\author{Benjamin Bayman and Shaul Hanany}%\fnref{footnote3}}
\address{University of Minnesota School of Physics and Astronomy, Minneapolis, MN 55455}
% \fntext[footnote3]{Additional information about the second and third authors}
\ead{hanany@umn.edu}

\author{Hugo Franco and Justin Marsh}%\fnref{footnote4}}
\address{Columbia Scientific Balloon Facility, Palestine, TX 75803}
% \fntext[footnote4]{Additional information about the co-authors}
\ead{hugo.franco@csbf.nasa.gov}

\author{Joy Didier and Amber D. Miller}%\fnref{footnote4}}
\address{Physics Department, Columbia University, New York, NY 10027}
% \fntext[footnote4]{Additional information about the co-authors}
\ead{amber@phys.columbia.edu }

\begin{abstract}

During the EBEX2013 long duration flight the payload was free to rotate in azimuth.
The observed azimuth motion consisted of a superposition of full rotations with a period of 10-30~minutes and oscillatory motion with an amplitude of tens of degrees, average period of 79~s, and period dispersion of 12~s.
We interpret the full rotations as induced by slow rotations of the balloon and the shorter period oscillatory motion as due to torsional oscillations of the flight line.
We derive the torsional stiffness of the flight line using the bifilar pendulum model and apply it to the flight line of the EBEX2013 payload.
We find a torsional spring constant of~36~kg~m$^2$/s$^2$ corresponding to a period of 58~s. 
We conclude that the bifilar model, which accounts for the geometry of the flight line but neglects all material properties, predicts a stiffness and period that are 45\% larger and 25\% shorter than those observed.
% k = 4 pi^2 I / T^2 = 4 pi^2 3100 / 58^2 = 19.6, (36-20)/36 = 44% -> 45%
% (79 - 58) / 79 = 26.6% -> 25%
It is useful to have a simple, easy to use, coarse approximation for the torsional constant of the flight line. 

\end{abstract}


\end{frontmatter}

\parindent=0.5 cm

\section{Introduction}
\label{sec:intro}

Stratospheric balloon payloads are suspended below a helium balloon by means of a flight line that typically consists 
of a parachute and cables. The dynamics of this system have been investigated in a number of 
publications~\citep{QJ:QJ49712757314, Yajima2009, 2009JSpRo..46..126M, Alexander2011736}.
To date, however, limited attention has been given to the free rotational motion of the system about the 
gravitational acceleration vector. Expressions for the torsional constant of the flight line have been 
given but nearly no measurements are available.~\citep{Morris1975, DUCARTERON1993185, TREILHOU20001423, FisselThesis}.

For nearly all payloads, the balloon has a moment of inertia that is orders of magnitude larger than that of  the payload.
Therefore, rotation of the balloon causes a rotation of the payload.
Superposed on this rotation is an oscillatory motion of the payload due to the torsional stiffness of the flight line.
It is useful to have a model with which to quantify the torsional stiffness of the flight line and therefore the 
characteristic frequency of oscillations. The characteristic rotational frequency represents 
a mechanical resonance of the system. With a validated model for the resonance one can predict the rotational 
motion of the payload, and, if necessary or desired, either use or effectively suppress the resonance to 
improve the performance of the payload. For example, a payload that requires rotational scanning requires least 
power if it excites the rotational resonance mode.

Unique opportunity to investigate the free torsional motion of a stratospheric payload was offered during the flight of the \ac{EBEX} experiment.
\ac{EBEX} was a balloon-borne polarimeter flown aboard a zero-pressure balloon that was launched on 
December 29, 2012 from McMurdo, Antarctica~\citep{joy_ieee_paper, Aubin_MGrossman2016, EBEXPaper3}.
The payload was designed to control attitude and scan a 400 sq.\ deg. area of the sky.
To realize the science goals, which were measurements of the polarization of the cosmic microwave background radiation, the collaboration implemented a set of pointing sensors to provide post-flight attitude determination accuracy of 15~arcsec RMS~\citep{EBEXPaper3}.
However, an error in the thermal design of the azimuth motor controller that was discovered when the payload reached float altitude rendered the azimuth motor inoperable.
For the rest of the flight the payload executed free motion in azimuth.
We use the combination of good  instantaneous attitude determination and free azimuth motion to quantify the rotational motion of the payload and to compare it to a model based on the bifilar approximation.
Specifically, we use the bifilar approximation to derive a rotational stiffness, also known as a rotational spring constant, due to the payload and flight line and compare it to the one derived from observed azimuthal oscillations during the EBEX flight. 

In Section~\ref{sec:azmotion} we give a qualitative description of the observed motion; section~\ref{sec:analysis} describes our analysis of the inflight attitude data.   
Section~\ref{sec:model} gives a derivation of the torsional stiffness of the flight line in the bifilar approximation.
In Section~\ref{sec:predictions} we use the approximation and the parameters of the EBEX payload to generate predictions for the period of torsional oscillations of the payload, and in Section~\ref{sec:discussion} we discuss our results. 
  % general intro of gondola motion
\section{Qualitative Azimuth Motion}
\label{sec:azmotion}

\begin{figure}[ht]
  \centering
  \includegraphics[width=0.9\textwidth]{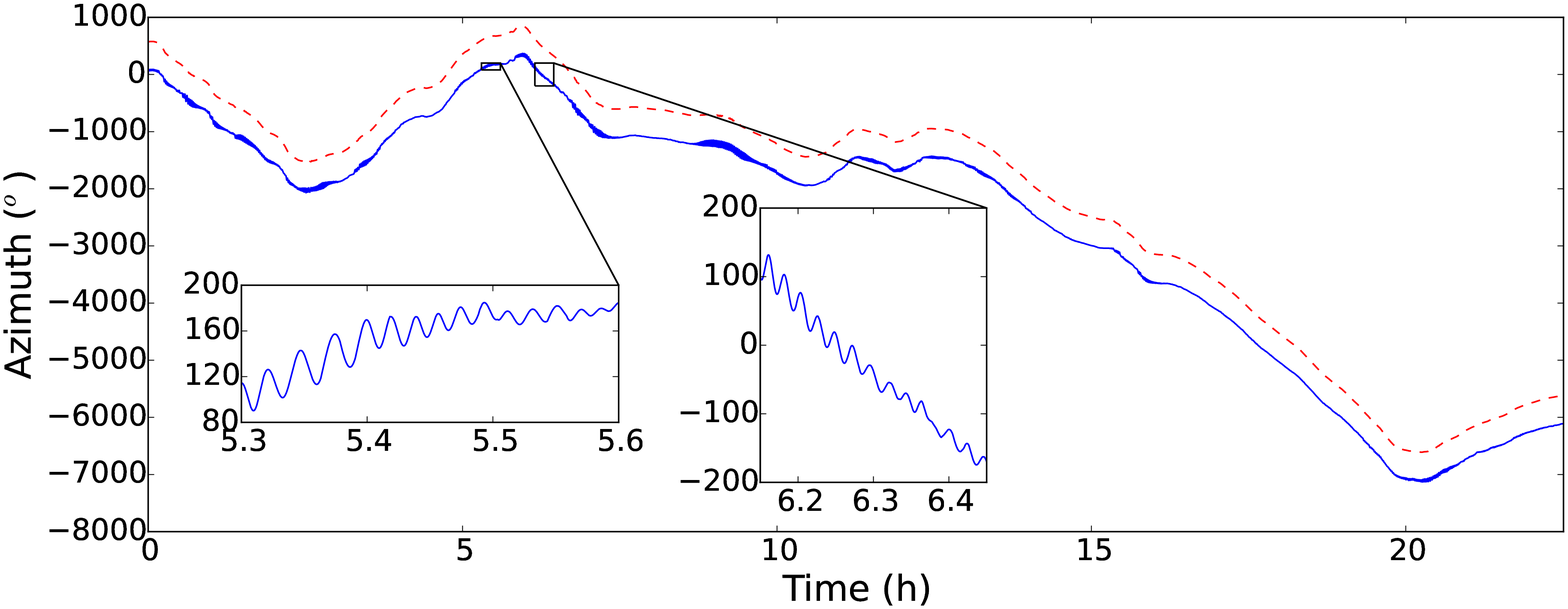}
  \caption{Azimuthal motion of the EBEX gondola during one 22.5 hour long segment of flight (solid) and 
    a moving average calculated with a 160~s window (dashed). 
    The moving average is offset by 500$^o$ for clarity.
    The motion is a superposition of slow continuous rotation and an azimuthal oscillation; see insets.
    Both the average rotation speed and amplitude of oscillations vary.
    They are 0.07$^0$/s (0.2$^0$/s) and up to 10$^o$ (20$^o$), in the left (right) inset.
  }
  \label{fig:azMotion}
\end{figure}

Figure~\ref{fig:azMotion} shows the azimuth motion of the \ac{EBEX} gondola 
during a representative segment of the flight.
The azimuth is defined so a value of zero corresponds to North.
Thereafter successive full rotations are summed to give cumulative azimuth and thus clearer 
indication of the motion of the payload.
For example, for the first 2.5~hours the payload rotates continuously in one direction; it then rotates 
in the other direction for approximately 3.5~hours.
We interpret this rotation as being induced by the rotation of the entire balloon.
Superposed on this slow rotation are azimuthal oscillations with a period of about 80~s, as shown by 
the insets in Figure~\ref{fig:azMotion}. These are due to the torsional constant of the flight line.
%In this paper we characterize the azimuthal oscillatory motion and model it as simple bifilar pendulum.
  % EBEX2013 flight, gondola motion
\section{Analysis}
\label{sec:analysis}

% moving mean, window=2*80=160 s
% subtract moving mean
% band-pass 10 order butterworth between 0.005 and 0.02 Hz (signal at 0.0125 Hz)
% diff + moving mean
% diff + moving mean
% min/max: |diff| < 1e-2 deg/s and diff2 < 0, keep one per consecutive

We characterize the azimuthal oscillatory motion by finding its time dependent period and amplitude.
We band-pass the data to leave only the oscillatory motion, find extrema, and from pairs of adjacent extrema extract 
periods and amplitudes. 

\begin{figure}[ht]
  \centering
  \includegraphics[width=0.45\textwidth]{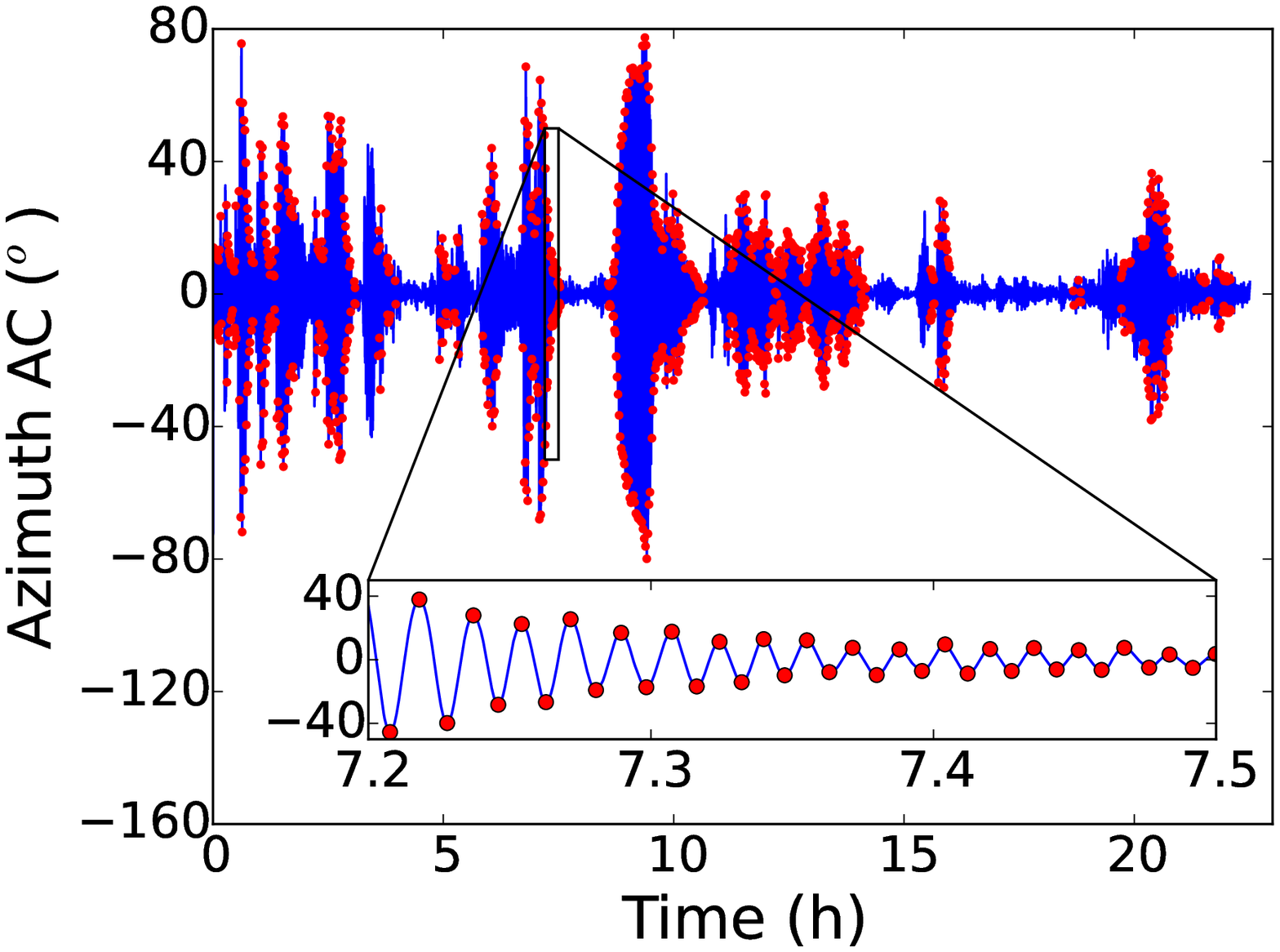}
  \includegraphics[width=0.45\textwidth]{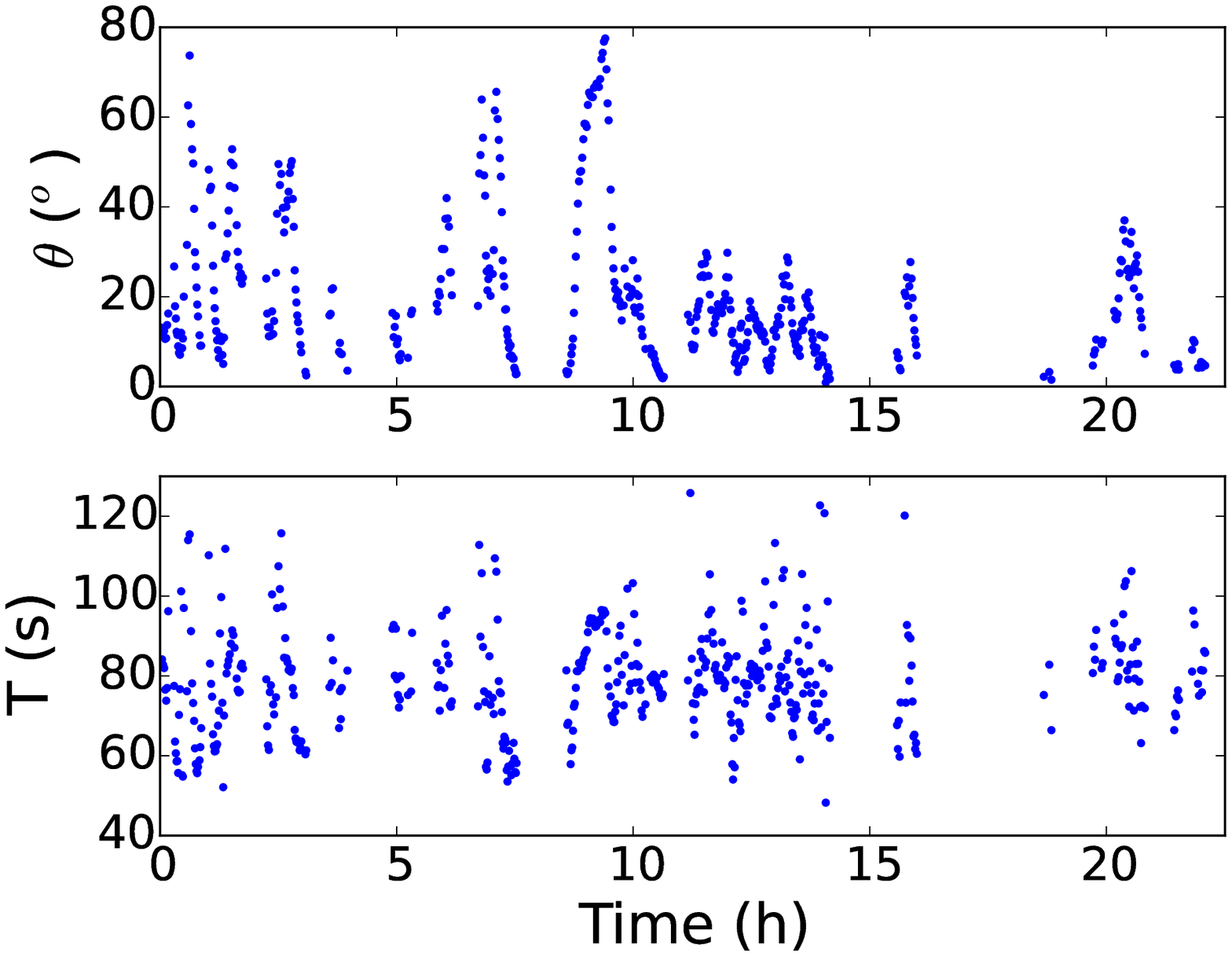}
  \caption{Offset-subtracted and band-passed azimuth time-ordered data during 22.5~hours segment (left).
    Red dots indicate extrema identified by the analysis. 
    The amplitude (top right) and period (bottom right) extracted from the oscillations identified on the left panel.
  }
  \label{fig:dataProcessing}
\end{figure}

We use the reconstructed attitude for two at-float flight segments that are in total 40 hours long.
A portion of the data is shown in Figure~\ref{fig:azMotion}.
To extract the oscillatory motion we subtract from the azimuth data an offset. The offset is a 
moving average of length 160~s, which is subsequently filtered with tenth-order Butterworth band-pass between 0.005 and 0.02~Hz.
%The band is selected to preserve and isolate the oscillatory motion.
The resulting data for a 22.5~hour segment is shown in the left panel of Figure~\ref{fig:dataProcessing}.
To find extrema we calculate the first and second time derivatives of the offset-subtracted data. The derivatives
are noisy and we therefore use a moving mean with a 10~s window. 
An extremum is defined to have a first derivative with an absolute value smaller than 0.01$^o$/s.
We only include in subsequent analysis pairs of consecutive local maxima and minima.
From each pair of extrema we extract a period $T$ and an amplitude $\theta$; see the right panel of Figure~\ref{fig:dataProcessing}. 
We estimate a 3~s and $0.07^{o}$ errors on a period and amplitude, respectively. 
%Our analysis does not identify all the oscillations in the data. It is particularly incomplete for small amplitude oscillations. 

\begin{figure}[ht]
  \centering
  \includegraphics[width=0.45\textwidth]{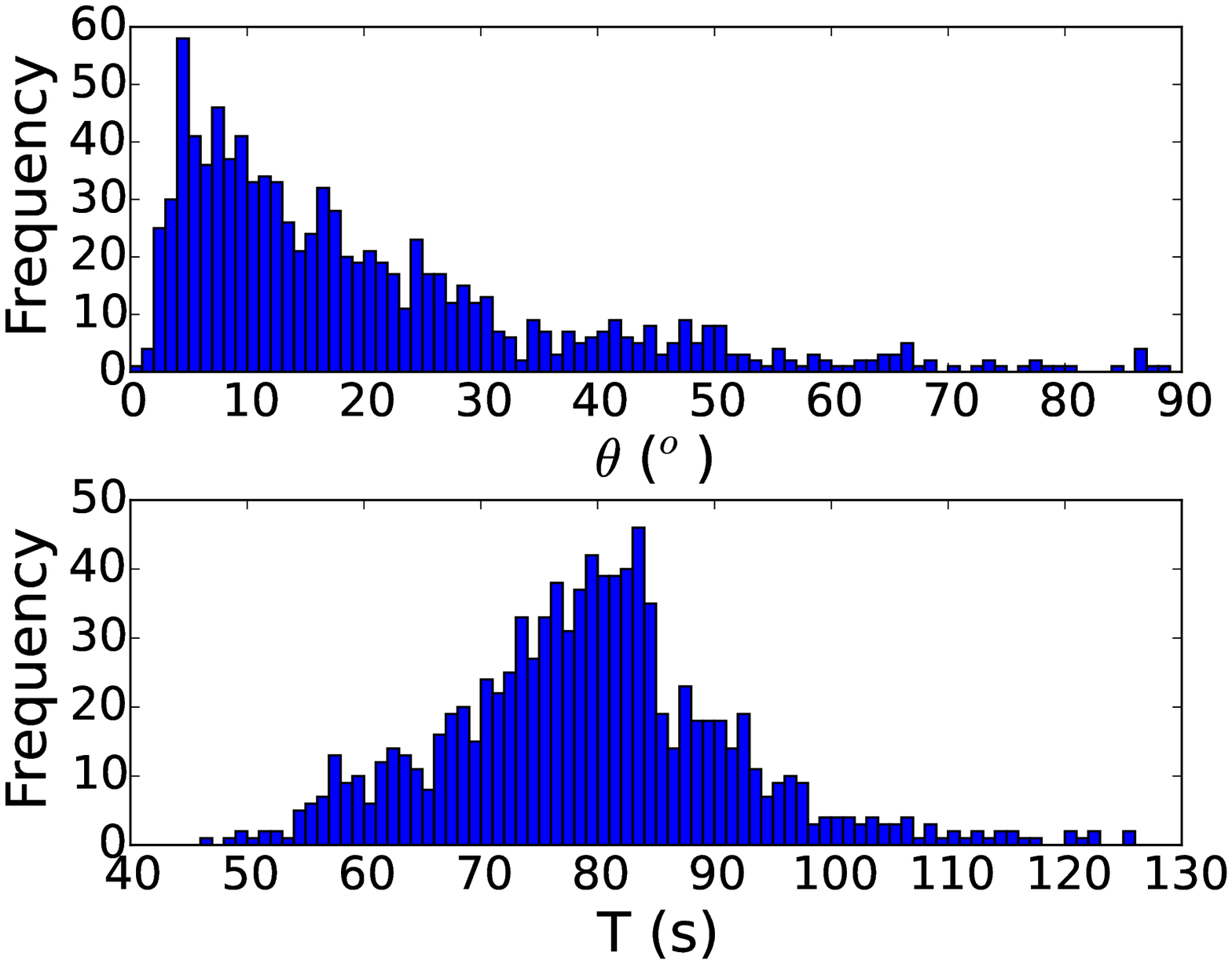}
  \includegraphics[width=0.45\textwidth]{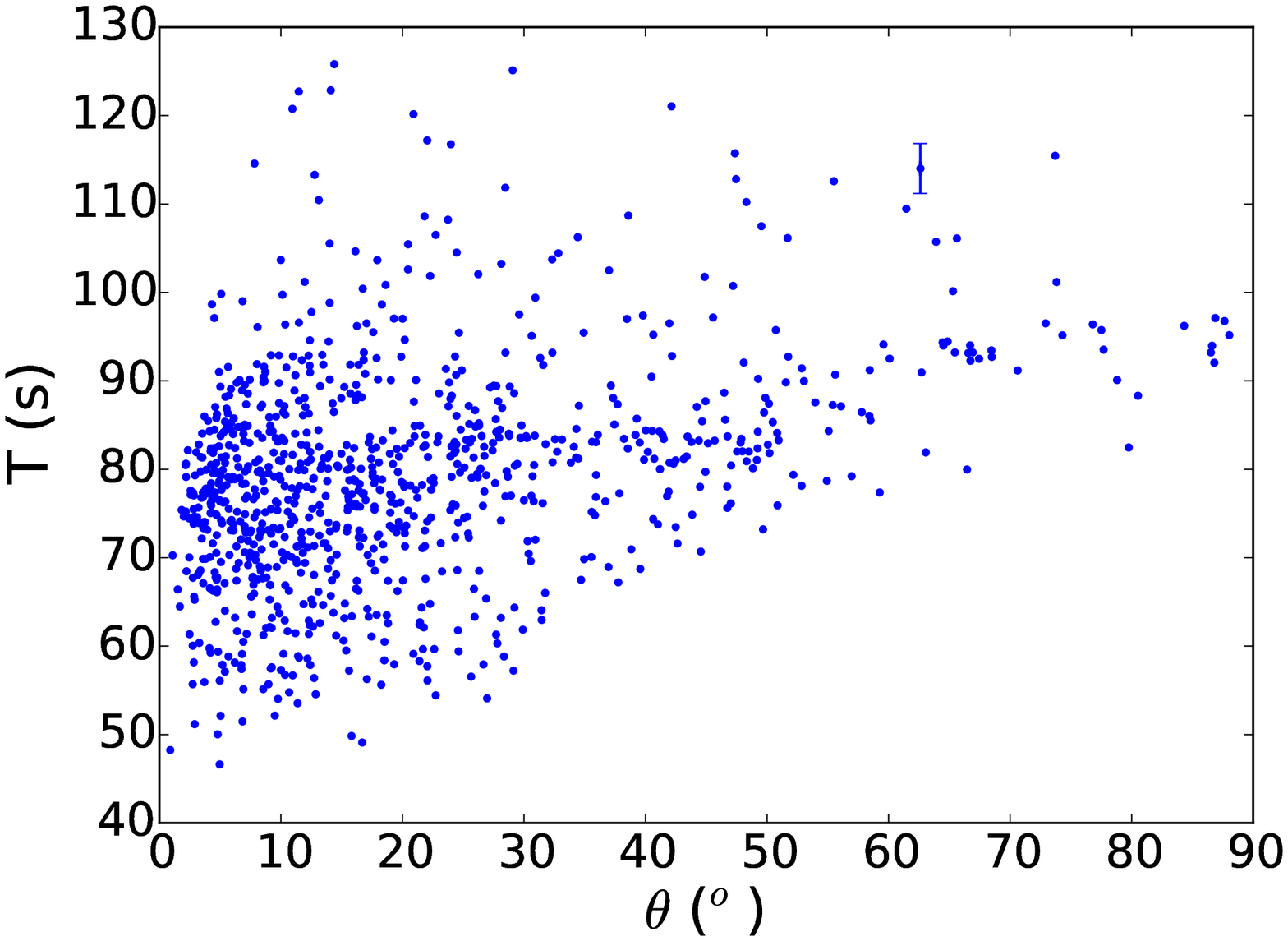}
  \caption{The distributions of measured amplitudes (top left) and periods (bottom left) and the correlation between them (right).
    For one datum we show the estimated errors in amplitude and period.
    }
  \label{fig:distributionsAndScatter}
\end{figure}

Figure~\ref{fig:distributionsAndScatter} shows histograms of the amplitudes and periods 
for the entire data set. 
%and the correlation between the two.
The average period is 79~s and the standard deviation is 12~s.
While some amplitudes can reach 90$^{o}$ degrees, 80\% of the amplitudes are below 30$^o$. 
We note that our sample of amplitudes and periods is not complete because the extrema identification 
algorithm is biased to reject low amplitude oscillations.
As we discuss later, this bias should not affect our results regarding the torsional constant of the flight line. 
The periods and amplitudes are correlated; see Figure~\ref{fig:distributionsAndScatter}.
The coefficient of correlation is 0.37, although this value is also subject to the incompleteness bias. 
  % data analysis, results
\section{Model}
\label{sec:model}

The flight train connects the 3200~kg \ac{EBEX} gondola to a 960,000~m$^{3}$ balloon.  
The moment of inertia of the payload is estimated to be 3100~kg~m$^{2}$.
We neglect moment of inertia of the flight train as it represents less than 2\% of the moment of inertia of the payload. 
% calculating sum(MR^2) for all the components, setting higher limit. Parachute is 44, total is 58 compare to 3100.
The moment of inertia of the inflated balloon is 3.3$\times$10$^6$~kg~m$^{2}$ and therefore the balloon is considered 
stationary relative to the azimuthal oscillations of the payload. 

\begin{figure}[ht]
  \centering
  \includegraphics[height=0.4\textheight]{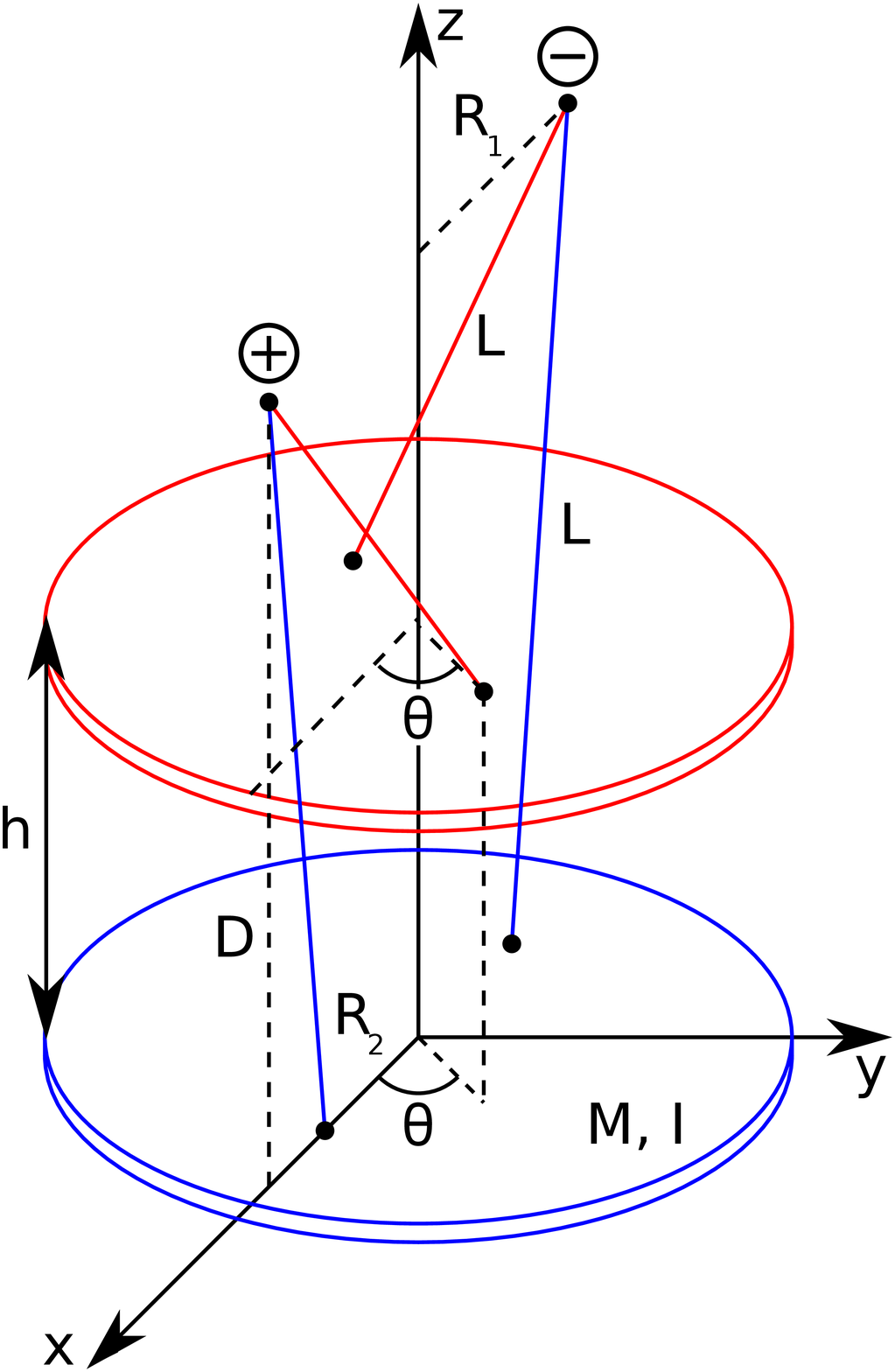}\hspace{1cm}
  \includegraphics[height=0.4\textheight]{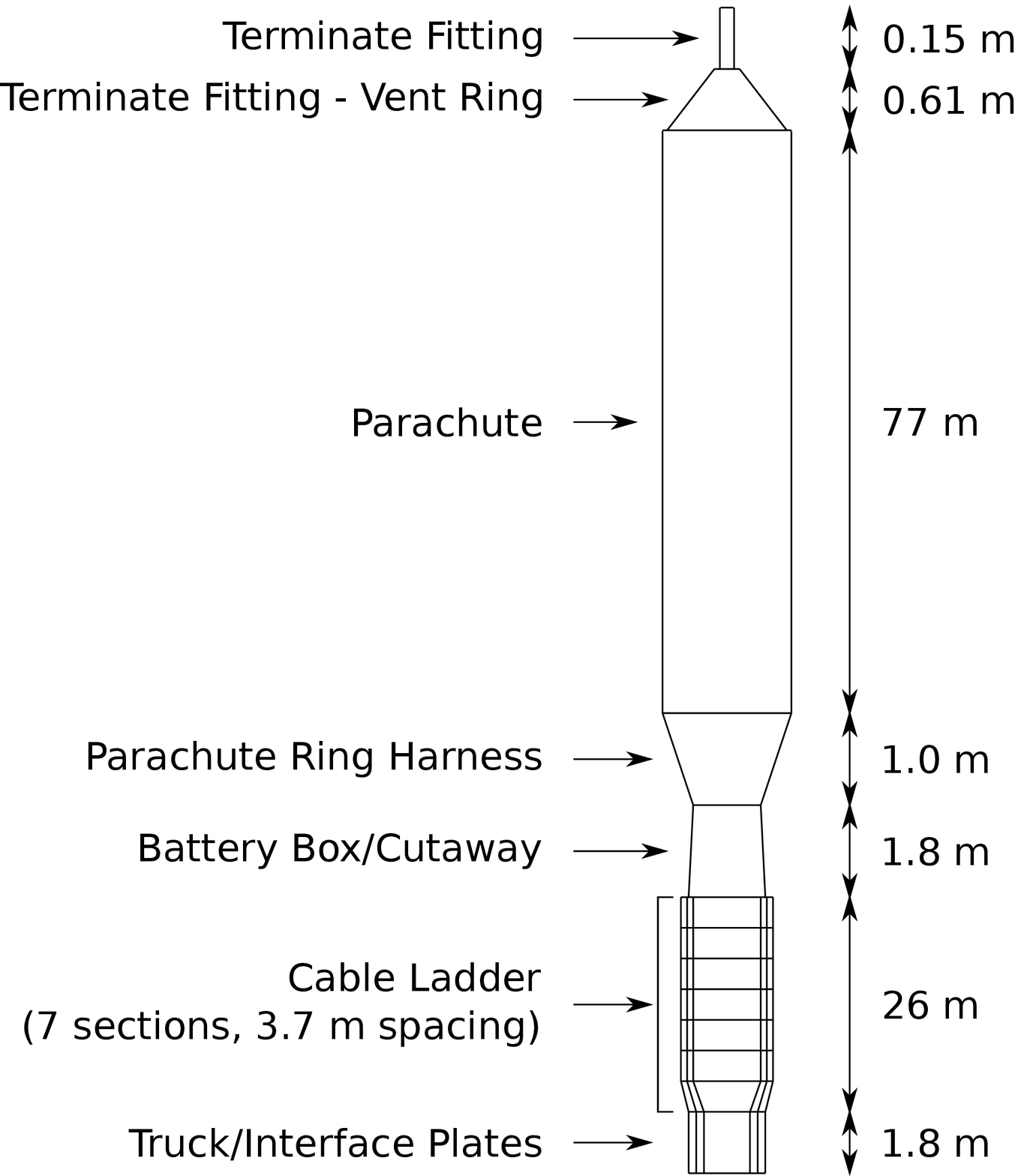}
  \caption{Left: Model of a bifilar pendulum.
    The equilibrium and non-equilibrium positions are shown in blue and red, respectively.
    % A mass $M$ with a moment of inertia $I$ is suspended at a vertical distance $D$ below the anchor points of two identical cables of length $L$.
    % The anchor points are separated by 2$R_1$ at the suspension and by 2$R_2$ on the load.
    % The rotation of the suspended mass by an angle $\theta$ (red) compared to the equilibrium position (blue) lifts the load by a height $h$ which increases the tension in the cables and in turn creates a restoring torque.
    Right: The flight train attaching the \ac{EBEX} gondola to the balloon.
    The parachute stretches to 77~m due to the payload.
    }
  \label{fig:flightTrainAndModel}
\end{figure}

We use the bifilar pendulum model to predict the period of rotational oscillations~\citep{MacMillan1936, Then1965}.
In this model a pair of massless cables of length $L$ suspend a mass $M$ with moment of inertia $I$, which is at a 
vertical distance $D$ below the anchor points. The anchor points are separated by 2$R_1$ at the suspension and by 2$R_2$ at the load.
As the mass rotates about the gravitational acceleration vector by an angle $\theta$ it rises by a height $h$, 
as shown in Figure~\ref{fig:flightTrainAndModel}. 
The restoring force is due to gravity. 
In this approximation the material properties of the suspension structure are irrelevant. The torsional stiffness $k$ is proportional 
to the mass $M$ and to the geometry of suspension. The period of oscillation is given by
\begin{equation}
T = 2 \pi \sqrt{\frac{I}{k}}.
\label{eq:period}
\end{equation}
Since the flight line is made up of several segments of different geometries we model it as a combination 
of rotational springs that add either in parallel or in series. 

%There are two equivalent ways to relate $k$ to the mass $M$ and to the geometry of suspension: force and energy.  They are both useful. 

%\subsection{Energy Approach} 
%\label{sec:modelEnergy}

We use energy considerations to derive the stiffness of the rotational spring. 
As the mass rotates by angle $\theta$ it rises a distance $h(\theta)$ and gains potential energy $U(h) = Mgh(\theta)$.
We calculate the increase in potential energy and relate the derivative of the potential energy to torque $\overrightarrow{\tau} (\theta)$.

%Equating the increase in gravitational potential energy to the energy stored in the 
%rotational spring $U(h) = Mgh = U(\theta) = \frac{1}{2} k \theta^2$ gives $k$ when the amplitude of the oscillations
%are small. 

At equilibrium the two anchor points at the top of the cables have coordinates $\pm R_1 {\bf {\hat x}} + D {\bf {\hat z}}$ and the two 
anchor points on the load have coordinates $\pm R_2 {\bf {\hat x}}$.
The cables have a length 
\begin{eqnarray}
L & = & |~ (\pm R_1 \mp R_2) {\bf {\hat x}} + D {\bf {\hat z}} ~| \nonumber \\
& = & \sqrt{(R_1 - R_2)^2 + D^2}.
\label{eq:Lequilibrium}
\end{eqnarray}
When the mass is rotated by $\theta$, the cables move out of the ${\bf {\hat x}}$-${\bf {\hat z}}$ plane but remain of length L.
This requires the lower end of the cable to rise a distance $h$.
The anchor points on the load then have coordinates $\pm R_2 \mathrm{cos} \theta {\bf {\hat x}} \pm R_2 \mathrm{sin} \theta {\bf {\hat y}} + h {\bf {\hat z}}$ and
\begin{eqnarray}
L & = & |~ \pm (R_1 - R_2 \textrm{cos} \theta) {\bf {\hat x}} \mp R_2 \textrm{sin} \theta {\bf {\hat y}} + (D - h) {\bf {\hat z}} ~| \nonumber \\
& = & \sqrt{(R_1-R_2 \mathrm{cos} \theta)^2 + (R_2 \mathrm{sin} \theta)^2 + (D-h)^2}.
\label{eq:LnonEquilibrium}
\end{eqnarray}
Combining equations~\ref{eq:Lequilibrium} and~\ref{eq:LnonEquilibrium} leads to a quadratic equation for $h$
% \begin{equation}
% 0 = h^2 - 2 h D + 4 R_1 R_2 \mathrm{sin}^2{\left( \frac{\theta}{2} \right)}
% \end{equation}
with one physical solution
\begin{equation}
h = D \left( 1 - \sqrt{1 - \frac{4 R_1 R_2}{D^2} \mathrm{sin}^2{\left( \frac{\theta}{2} \right)} } \right).
\label{eq:bifilarRsise}
\end{equation}
Using $| \overrightarrow{\tau} (\theta) | = - d U (\theta) / d \theta$ we find
\begin{equation}
| \overrightarrow{\tau} (\theta) | = \frac{- M g R_1 R_2 \mathrm{sin} \theta}{\sqrt{L^2 - R_1^2 - R_2^2 + 2 R_1 R_2 \mathrm{cos} \theta}}; 
\label{eq:torque}
\end{equation}
\citet{Morris1975} gives the same expression.
The torsional stiffness $k$ is 
\begin{equation}
k = - \frac{d | \overrightarrow{\tau} (\theta) |}{d \theta},
\end{equation}
leading to
\begin{equation}
k = \frac{M g R_1 R_2}{\sqrt{L^2 - R_1^2 - R_2^2 + 2 R_1 R_2 \mathrm{cos} \theta}} \left[ \mathrm{cos} \theta + 
\frac{R_1 R_2 \mathrm{sin}^2 \theta}{L^2 - R_1^2 - R_2^2 + 2 R_1 R_2 \mathrm{cos} \theta} \right],
\label{eq:kLargeAngles}
\end{equation}
The period can be calculated using Equation~\ref{eq:period}. 

For small amplitudes of oscillation, denoted $\theta_{s}$, and a long pendulum ($D \approx L \gg R_1, R_2$) we 
find 
\begin{eqnarray}
h_{\theta_s} = \frac{R_1 R_2 \theta_s^2}{2 L}.
\end{eqnarray}
In this approximation we can set $U(h_{\theta_s}) = Mgh_{\theta_s} = U(\theta_s) = \frac{1}{2} k_{\theta_s} \theta_s^2$ to find
\begin{eqnarray}
k_{\theta_s} =\frac{M g R_1 R_2}{L},
\label{eq:kThetaSmallerThanOnesPotentialEnergy}
\end{eqnarray}
which is the small angle limit of Equation~\ref{eq:kLargeAngles} (see also~\citet{Morris1975}).
For small angles the period is 
\begin{equation}
T_{\theta_s} = 2\pi \sqrt{ \frac{I L}{ {M g R_1 R_2} } }.
\label{eq:TthetaSmallerThanOnes}
\end{equation}

  % bifilar approximation
\section{Model Predictions}
\label{sec:predictions}

Several components that are attached in series constitute the flight train; see Figure~\ref{fig:flightTrainAndModel} 
and Table~\ref{tab:flightTrainComponents}.
We model each component as a bifilar pendulum. Each component may itself be composed of a number of 
bifilar pendula connected in parallel. 
This represents a simplification of the real components in terms of their geometry, and it neglects material 
properties that affect the torsional stiffness. 
The mass and the dimensions of each component were measured on the ground. The equivalent bifilar 
dimensions are given in Table~\ref{tab:flightTrainComponents}.

The terminate fitting and the battery box/cutaway are both solid cylinders; the terminate fitting vent ring is a truncated cone.  They are all simulated as simple bifilar pendula with their actual length and only the outer radii. 
The parachute has 130 parallel gores at equal radius. They are equivalent to a single bifilar pendulum at the same radius.
We use parachute's stretched length as measured on the ground when subjected to the same load as the suspended weight.
The parachute ring harness consists of 12 pairs of 3/16'' parallel steel cables.
It is equivalent to a single bifilar pendulum of the same radius.
The cable ladder is composed of three sets of parallel cables arranged in increasing radius.
They are all listed in Table~\ref{tab:flightTrainComponents} but when we calculate stiffness we only include the pair at the outermost radius.
Upon torsional motion that pair of cables carries the entire suspended load; the inner cables become slack.   
The truck and interface plates are solid, but are simulated as three parallel bifilar pendula with successively larger radius 
that all share the load.

The spring constant of the flight train is $k_{FT}^{-1} = \sum_{i} k_i^{-1}$, where $k_{i}$ is the spring constant of element $i$.
For the cases in which element $i$ is made of $N$ parallel pendula, the spring constant is $k_{i} = \sum_{j}^{N} k_{ij}$, where $k_{ij}$ is the spring constant of element $j$.
%For the cases in which element $i$ is made of $N$ parallel cables of different radii, we use the spring constant of the outermost element since it carries all the load. From Equation~\ref{eq:bifilarRsise}, we see that the parallel element with the biggest product $R_1 R_2$ lifts the whole suspended mass. For instance, the outer layer of the cable ladder 1 lifts the load by 3.3~mm as the inner layer would lift it by only 1.8~mm at an angle of 30$^0$.
Table~\ref{tab:flightTrainComponents} also gives the calculated spring constants using Equation~\ref{eq:kThetaSmallerThanOnesPotentialEnergy}.
For each entry in the table, the mass assumed is the mass suspended below that element. 
The equivalent spring constant of the flight train is 36~kg~m$^2$ giving a small oscillation period of 58~s. 
We conservatively estimate the error on the moment of inertia and mass of the gondola, and the length and radii of each 
elements of the flight train to be 10\%.
These errors are uncorrelated and with standard error propagation we find a 7\% (4~s) error on the predicted period.

\begin{table}[ht!]
\caption{Physical characteristics of the flight train components modeled as bifilar pendula. 
Rows separated by lines represent sequential elements of the flight train for which the spring constants add like resistors in parallel.
For the cable ladder we show its geometry, but only the outermost pair of carry the load during rotations. 
The truck and interface plates are solid and simulated as three successive parallel pairs of cables.}
\centering
\footnotesize
\begin{tabular}{| l | c |  c | c |  c | c | c | c | c | c |}\hline
Flight Train Element & Mass & $M$ & $R_1$ & $R_2$ & $L$ & $k_{\theta_s ij}$ &$k_{\theta_s i}$\\
& kg & kg & m & m & m & $\frac{\mathrm{kg}~\mathrm{m}^2}{\mathrm{s}^2}$ & $\frac{\mathrm{kg}~\mathrm{m}^2}{\mathrm{s}^2}$\\ \hline \hline
Terminate Fitting                     & 0.91  &  3634 & 0.046 & 0.046 & 0.15  & 490   & 490  \\ \hline
Terminate Fitting - Vent Ring         & 2.3   &  3632 & 0.082 & 0.39  & 0.61  & 1900  & 1900 \\ \hline
Parachute                             & 250   &  3381 & 0.42  & 0.42  & 77    & 76    & 76   \\ \hline
Parachute Ring Harness                & 18    &  3363 & 0.42  & 0.22  & 1.0   & 3000  & 3000 \\ \hline
Battery Box/Cutaway                   & 72    &  3290 & 0.22  & 0.25  & 1.8   & 1000  & 1000 \\ \hline
Cable Ladder 1, Outer                 & 3.5   &  3280 & 0.30  & 0.30  & 3.7   & 790   &      \\
Cable Ladder 1, Middle                & 3.5   &     0 & 0.26  & 0.26  & 3.7   & 0     & 790  \\
Cable Ladder 1, Inner                 & 3.5   &     0 & 0.22  & 0.22  & 3.7   & 0     &      \\\hline
Cable Ladder 2, Outer                 & 3.5   &  3269 & 0.30  & 0.30  & 3.7   & 790   &      \\
Cable Ladder 2, Middle                & 3.5   &     0 & 0.26  & 0.26  & 3.7   & 0     & 790  \\
Cable Ladder 2, Inner                 & 3.5   &     0 & 0.22  & 0.22  & 3.7   & 0     &      \\\hline
Cable Ladder 3, Outer                 & 3.5   &  3259 & 0.30  & 0.30  & 3.7   & 790   &      \\
Cable Ladder 3, Middle                & 3.5   &     0 & 0.26  & 0.26  & 3.7   & 0     & 790  \\
Cable Ladder 3, Inner                 & 3.5   &     0 & 0.22  & 0.22  & 3.7   & 0     &      \\\hline
Cable Ladder 4, Outer                 & 3.5   &  3248 & 0.30  & 0.30  & 3.7   & 790   &      \\
Cable Ladder 4, Middle                & 3.5   &     0 & 0.26  & 0.26  & 3.7   & 0     & 790  \\
Cable Ladder 4, Inner                 & 3.5   &     0 & 0.22  & 0.22  & 3.7   & 0     &      \\\hline
Cable Ladder 5, Outer                 & 3.5   &  3238 & 0.30  & 0.30  & 3.7   & 780   &      \\
Cable Ladder 5, Middle                & 3.5   &     0 & 0.26  & 0.26  & 3.7   & 0     & 780  \\
Cable Ladder 5, Inner                 & 3.5   &     0 & 0.22  & 0.22  & 3.7   & 0     &      \\\hline
Cable Ladder 6, Outer                 & 3.5   &  3227 & 0.30  & 0.30  & 3.7   & 780   &      \\
Cable Ladder 6, Middle                & 3.5   &     0 & 0.26  & 0.26  & 3.7   & 0     & 780  \\
Cable Ladder 6, Inner                 & 3.5   &     0 & 0.22  & 0.22  & 3.7   & 0     &      \\\hline
Cable Ladder 7, Outer                 & 3.5   &  3216 & 0.30  & 0.25  & 3.7   & 650   &      \\
Cable Ladder 7, Middle                & 3.5   &     0 & 0.26  & 0.20  & 3.7   & 0     & 650  \\
Cable Ladder 7, Inner                 & 3.5   &     0 & 0.22  & 0.15  & 3.7   & 0     &      \\\hline
Truck/Interface Plates, Outer          & 5.4   &  1067 & 0.25  & 0.25  & 1.8   & 370   &      \\
Truck/Interface Plates, Middle         & 5.4   &  1067 & 0.20  & 0.20  & 1.8   & 240   & 740  \\
Truck/Interface Plates, Inner          & 5.4   &  1067 & 0.15  & 0.15  & 1.8   & 130   &      \\\hline \hline
Total                                 & 430   &  3200 & N/A   & N/A   & 110   & N/A   & 36    \\ \hline
\end{tabular}
\label{tab:flightTrainComponents}
\end{table}

  % bifilar approximation predictions
\section{Discussion}
\label{sec:discussion}

%In the previous section, we assumed small oscillations and a massless flight train.
%We discuss in this section the magnitude of these effects to validate the choice of the bifilar model.
%We also discuss the error propagation on the period to quantify the accuracy of the model.

%\subsection{Flight Train with Mass}

\begin{figure}[ht]
  \centering
  \includegraphics[width=0.45\textwidth]{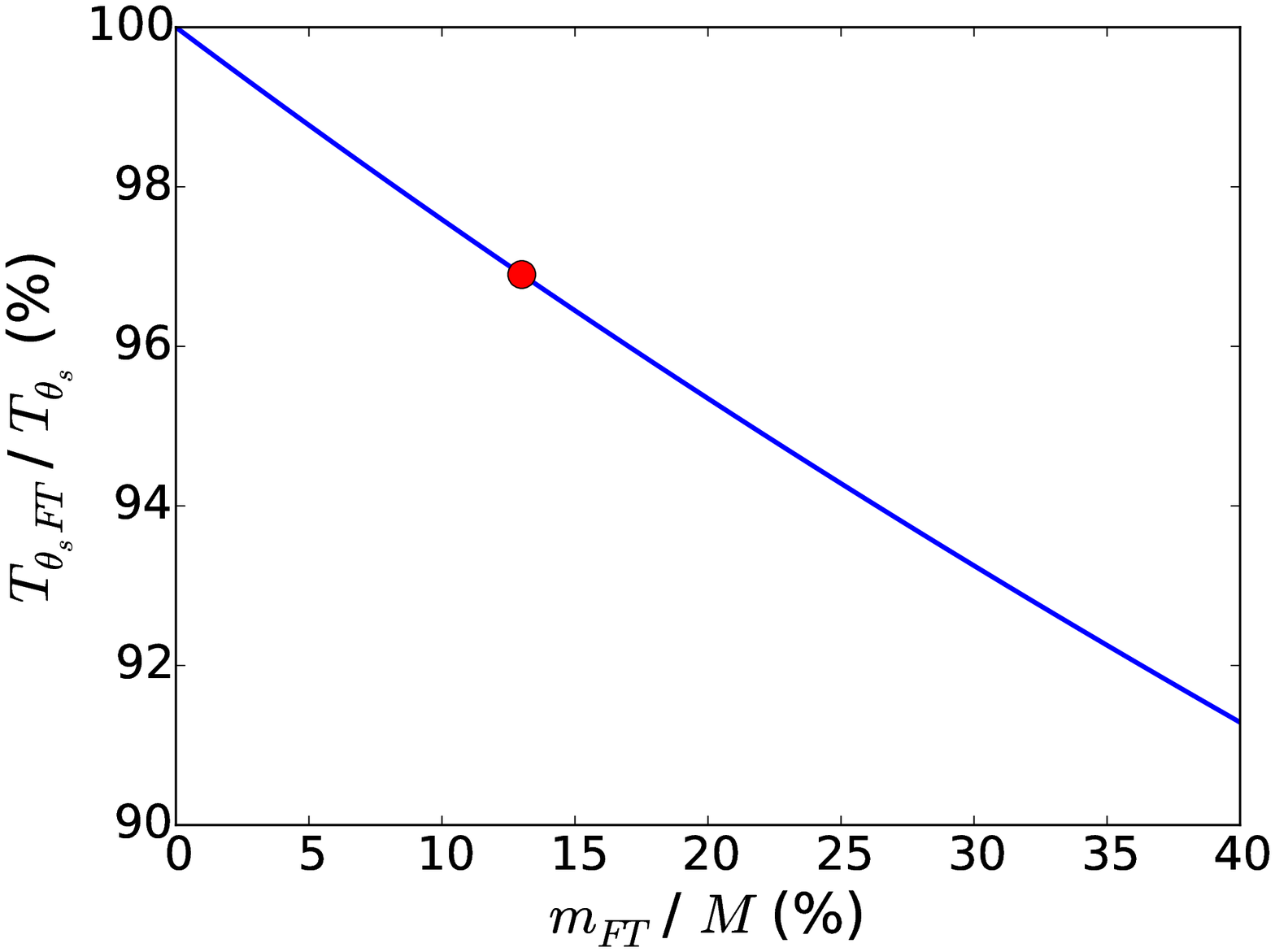}
  \includegraphics[width=0.45\textwidth]{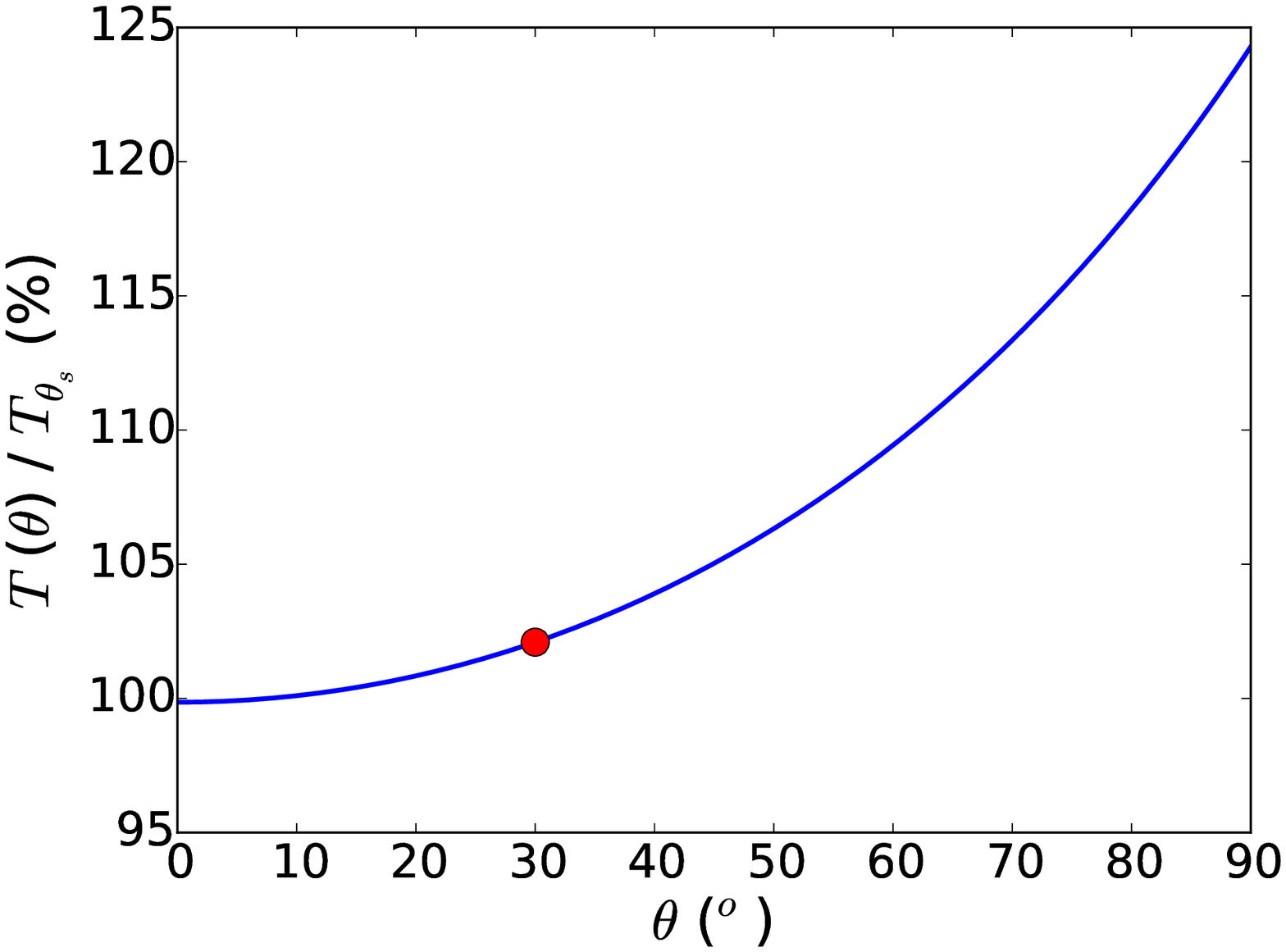}
  \caption{Left: Fractional change in period when assuming that the mass of the flight train is distributed uniformly along the length. 
    For our case the period decreases by 3\%.
    Right: Fractional change in period when calculating average oscillation periods as 
    a function of oscillation amplitude. For the case of 30$^{o}$ amplitude the period increases by 2\%.
  }
  \label{fig:Tcorrections}
\end{figure}

In the bifilar approximation the ropes are massless; the entire mass is concentrated in the suspended mass. 
However, in the EBEX case the mass of the flight train is 430~kg and is 13\% of the mass suspended below the 
truck and interface plates. % 430/3200=13.4%
%We assumed in Section~\ref{sec:model} that each of the components of the flight train is modeled as a bifilar 
% pendulum with massless ropes and with a mass equivalent to all of the components below suspended at its bottom. 
We investigate how the bifilar approximation changes when the flight train mass $m_{FT}$ is spread uniformly along the length.
When the suspended mass rises by $h_{\theta_s}$, the center of mass of the flight train rises by $h_{\theta_s}/2$ 
contributing to an extra potential energy contribution of $m_{FT}gh_{\theta_s}/2$ in the derivation 
of Equation~\ref{eq:kThetaSmallerThanOnesPotentialEnergy}.
Including the potential energy of the flight train modifies Equation~\ref{eq:TthetaSmallerThanOnes}
\begin{equation}
T_{\theta_sFT} \approx 2 \pi \sqrt{ \frac{I L}{M g R_1 R_2 \left( 1 + \frac{m_{FT}}{2M}\right)} }.
\label{eq:TflightTrainMass}
\end{equation}
For the case of the EBEX payload this correction increases the stiffness by 6\% and 
%The mass of the flight train is 620~kg and represents 19\% of the mass of the \ac{EBEX} gondola.
decreases the predicted period by 3\% (2~s) as shown in Figure~\ref{fig:Tcorrections}.
% stiffness is twice the period (relative) in the opposite direction
% this plot assumes that the mass of the flight train (430 kg) is distributed uniformly across the flight train. This is a worse case scenario compared to calculating it element by element:
% 1./(1./1.01 + 1./1.01) -> 1%
% M is roughly constant for the element of the FT, the parachute dominates with a contribution of 1.8%, this graph shows 3.2%.

Our analysis in Section~\ref{sec:model} relied on the small angle approximation. A significant fraction of the amplitudes
exceed small angles. As shown in the general case, Equation~\ref{eq:kLargeAngles},  
the bifilar spring stiffness depends on the oscillating angle and therefore 
changes as the gondola rotates.  
Remembering that $k = - d | \overrightarrow{\tau} (\theta') | / d\theta'$, we calculate an average spring constant using
\begin{equation}
\overline{k_{ij}} (\theta) = \frac{1}{2 \theta} \int_{-\theta}^{\theta} k_{ij}(\theta') d \theta' = \frac{| \overrightarrow{\tau_{ij}} (\theta) |}{\theta}
\label{eq:k2ndOrder}
\end{equation}
for every component of the flight train.
We then repeat the calculation in Section~\ref{sec:predictions} and the period is derived using $T (\theta) = 2 \pi \sqrt{I / \bar{k} (\theta)}$.
Figure~\ref{fig:Tcorrections} gives the ratio of the period calculated using the average stiffness to that calculated using the small 
angle approximation. The correction increases the period by 2\% (1~s) for amplitudes of 30$^o$.
%, which includes at least 80\% of the analyzed data during the EBEX2013 flight.

Our data analysis is biased toward selecting torsional periods with amplitudes that are larger than few degrees; see the discussion in 
Section~\ref{sec:analysis}. While the correlation plot shown in Figure~\ref{fig:distributionsAndScatter} indicates a correlation 
between amplitude and period, it also strongly indicates that inclusion of the smallest periods of oscillation would not alter 
the conclusion of mean period near 80~s by a significant fraction of the 12~s quoted dispersion. 

Our application of the bifilar model gives a period prediction that is lower by 25\% compared to the 
measured mean period of torsional oscillations.
The model predicts a stiffness of 36~kg~m$^2$/s$^2$; the measured mean period suggests an effective 
stiffness of 20~kg~m$^2$/s$^2$. % 4*pi^2 * 3100 / 79^2 = 19.6
It is instructive to learn that the simple, easy to apply bifilar model gives predictions of the stiffness and period that 
are within 45\% and 25\%, respectively of the measured values.
Using this simple model, payload designers can coarsely evaluate the expected modes of 
azimuthal oscillations and, if necessary, implement mechanisms to dampen the oscillatory resonance or use 
it as part of the motion control. It is also instructive to learn that the model gives a larger stiffness than the one observed.
A more accurate model, which is beyond the scope of this paper, would account more rigorously for the coupled nature 
of the torsional oscillators, for their material properties, for the non-uniform mass distribution of the flight train, 
and for the coupling of the rotational resonance mode to the vertical and pendulum oscillation modes.
\section{Acknowledgements}

Support for the development and flight of the EBEX instrument was provided by NASA grants NNX12AD50G, NNX13AE49G,
NNX08AG40G, and NNG05GE62G, and by NSF grants AST-0705134 and ANT-0944513.
We acknowledge support from the Italian INFN INDARK Initiative.
% Ade and Tucker acknowledge the Science \& Technology Facilities Council for its continued support of the underpinning technology for filter and waveplate development.
We also acknowledge support by the Canada Space Agency, the Canada Research Chairs Program, the Natural Sciences and
Engineering Research Council of Canada, the Canadian Institute for Advanced Research, the Minnesota Supercomputing Institute, the National Energy Research Scientific Computing Center, the Minnesota and Rhode Island Space Grant Consortia, our collaborating institutions, and Sigma Xi the Scientific Research Society.
% Baccigalupi acknowledges support from the RADIOFOREGROUNDS grant of the European Union's Horizon 2020 research and innovation program (COMPET-05-2015, grant agreement number 687312).
Didier acknowledges a NASA NESSF fellowship NNX11AL15H
% Reichborn-Kjennerud acknowledges an NSF Post-Doctoral          Fellowship AST-1102774, and a NASA Graduate Student Research Fellowship. Raach and Zilic acknowledge support by the Minnesota Space Grant       Consortium.
We very much thank Danny Ball and his colleagues at the Columbia Scientific Balloon Facility for their dedicated support of the EBEX program.
% We acknowledge the design work by Lorenzo Pascale.

\begin{acronym}
    %A
    \acro{ACS}{attitude control system}
    \acro{ADC}{analog-to-digital converters}
    \acro{ADS}{attitude determination software}
    \acro{AHWP}{achromatic half-wave plate}
    \acro{AMC}{Advanced Motion Controls}
    \acro{ARC}{anti-reflection coating}
    \acro{ATA}{advanced technology attachment}
    %B
    \acro{BRC}{bolometer readout crates}
    \acro{BLAST}{Balloon-borne Large-Aperture Submillimeter Telescope}
    %C
    \acro{CANbus}{controller area network bus}
    \acro{CMB}{cosmic microwave background}
    \acro{CMM}{coordinate measurement machine}
    \acro{CSBF}{Columbia Scientific Balloon Facility}
    \acro{CCD}{charge coupled device}
    %D
    \acro{DAC}{digital-to-analog converters}
    \acro{DASI}{Degree~Angular~Scale~Interferometer}
    \acro{dGPS}{differential global positioning system}
    \acro{DfMUX}{digital~frequency~domain~multiplexer}
    \acro{DLFOV}{diffraction limited field of view}
    \acro{DSP}{digital signal processing}
    %E
    \acro{EBEX}{E~and~B~Experiment}
    \acro{EBEX2013}{EBEX2013}
    \acro{ELIS}{EBEX low inductance striplines}
    \acro{EP1}{EBEX Paper 1}
    \acro{EP2}{EBEX Paper 2}
    \acro{EP3}{EBEX Paper 3}
    \acro{ETC}{EBEX test cryostat}
    %F
    \acro{FDM}{frequency domain multiplexing}
    \acro{FPGA}{field programmable gate array}
    \acro{FCP}{flight control program}
    \acro{FOV}{field of view}
    \acro{FWHM}{full width half maximum}
    %G
    \acro{GPS}{global positioning system}
    %H
    \acro{HPE}{high-pass edge}
    \acro{HWP}{half-wave plate}
    %I
    \acro{IA}{integrated attitude}
    \acro{IP}{instrumental polarization} 
    %J
    \acro{JSON}{JavaScript Object Notation}
    %L
    \acro{LDB}{long duration balloon}
    \acro{LED}{light emitting diode}
    \acro{LCS}{liquid cooling system}
    \acro{LC}{inductor and capacitor}
    \acro{LPE}{low-pass edge}
    %M
    \acro{MLR}{multilayer reflective}
    \acro{MAXIMA}{Millimeter~Anisotropy~eXperiment~IMaging~Array}
    %N
    \acro{NASA}{National Aeronautics and Space Administration}
    \acro{NDF}{neutral density filter}
    %P
    \acro{PCB}{printed circuit board}
    \acro{PE}{polyethylene}
    \acro{PTFE}{polytetrafluoroethylene}
    \acro{PME}{polarization modulation efficiency}
    \acro{PSF}{point spread function}
    \acro{PV}{pressure vessel}
    \acro{PWM}{pulse width modulation}
    %R
    \acro{RMS}{root mean square}
    %S
    \acro{SLR}{single layer reflective}
    \acro{SMB}{superconducting magnetic bearing}
    \acro{SQUID}{superconducting quantum interference device}
    \acro{SQL}{structured query language}
    \acro{STARS}{star tracking attitude reconstruction software}
    %T
    \acro{TES}{transition edge sensor}
    \acro{TDRSS}{tracking and data relay satellites}
   \acro{TM}{transformation matrix}

\end{acronym}

%%%%%%%%%%%%%%%%%%%%%%%%%%%%%%%%%%%%%%%%%%%%%%%%%%%%%%%%%%%%%%%%%%%%%%%%%%%%%
%% Appendices
% \appendix


\begin{thebibliography}{}

% \bibitem[Names(Year)]{label} or \bibitem[Names(Year)Long names]{label}.
% (\harvarditem{Name}{Year}{label} is also supported.)

\bibitem[Alexander \& de la Torre(2011)]{Alexander2011736}
Alexander, P. \& de la Torre, A. 2011,
Uncertainties in the measurement of the atmospheric velocity due to balloon-gondola pendulum-like motions,
Advances in Space Research, 47, 4, 736-739.

\bibitem[Aubin et al.(2016)]{Aubin_MGrossman2016}
Aubin, F., Aboobaker, A.M., Ade, P., Araujo, D., Baccigalupi, C., Bao, C., Borrill, J., Chapman, D., Didier, J., Dobbs, M., Feeney, S., Geach, C., Hanany, S., Helson, K., Hillbrand, S., Hilton, G., Hubmayr, J., Jaffe, A., Johnson, B., Jones, T., Kisner, T., Klein, J., Korotkov, A., Lee, A., Levinson, L., Limon, M., Macdermid, K., Marchenko, V., Miller, A.D., Milligan, M., Pascale, E., Puglisi, G., Raach, K., Reichborn-Kjennerud, B., Reintsema, C., Sagiv, I., Smecher, G., Stompor, R., Tristram, M., Tucker, G.S., Westbrook, B., Young, K., \& Zilic, K. 2016,
Temperature calibration of the E and B experiment,
Proceedings of the Fourteenth Marcel Grossman Meeting on General Relativity (in press).

\bibitem[Didier et al.(2015)]{joy_ieee_paper}
Didier, J., Chapman, D., Aboobaker, A.A., Araujo, D., Grainger, W., Hanany, S., Helson, K., Hillbrand, S., Korotkov, A., Limon, M., Miller, A., Reichborn-Kjennerud, B., Sagiv, I., Tucker, G. \& Vinokurov, Y. 2015,
A high-resolution pointing system for fast scanning platforms: The EBEX example,
Proc. of IEEE Aerospace Conference.

\bibitem[Ducarteron \& Treilhou(1993)]{DUCARTERON1993185}
Ducarteron, J.P. \& Treilhou, J.P. 1993,
Resonance frequencies of a gondola submitted to a forced rotation under a stratospheric balloon
Advances in Space Research, 13, 2, 185-188.

\bibitem[Dvorkin et al.(2001)]{QJ:QJ49712757314}
Dvorkin, Y., Paldor, N. \& Basdevant, C. 2001,
Reconstructing balloon trajectories in the tropical stratosphere with a hybrid model using analysed fields,
Quarterly Journal of the Royal Meteorological Society, 127, 753, 975-988.

\bibitem[Fissel(2013)]{FisselThesis}
Fissel, L.M. 2013,
Probing the Role Played by Magnetic Fields in Star Formation with BLASTPol
University of Toronto.

% \bibitem[Klopsteg(1930)]{Klopsteg1930}
% Klopsteg, P.E. 1930,
% The Bifilar Pendulum,
% Review of Scientific Instruments, 1, 3-8.

\bibitem[MacMillan(1936)]{MacMillan1936}
MacMillan, W.D. 1936,
Dynamics of Rigid Bodies,
McGraw-Hill, New York, 130-133.

\bibitem[Morani et al.(2009)]{2009JSpRo..46..126M}
Morani, G., Palumbo, R., Cuciniello, G., Corraro, F. \& Russo, M. 2009,
Method for Prediction and Optimization of a Stratospheric Balloon Ascent Trajectory,
Journal of Spacecraft and Rockets, 46, 126-133.

\bibitem[Morris(1975)]{Morris1975}
Morris, A.L. 1975,
Scientific Ballooning Handbook,
NCAR Technical Note, NCAR-TN/IA-99, National Center for Atmospheric Research.

\bibitem[The EBEX Collaboration(2017)]{EBEXPaper3}
The EBEX Collaboration, Aboobaker, A.M., Ade, P., Araujo, D., Aubin, F., Baccigalupi, C., Bao, C., Chapman, D., Didier, J., Dobbs, M., Grainger, W., Hanany, S., Helson, K., Hillbrand, S., Hubmayr, J., Jaffe, A., Johnson, B., Jones, T., Klein, J., Korotkov, A., Lee, A., Levinson, L., Limon, M., MacDermid, K., Miller, A.D., Milligan, M., Moncelsi, L., Pascale, E., Raach, K., Reichborn-Kjennerud, B., Sagiv, I., Tucker, C., Tucker, G.S., Westbrook, B., Young, K. \& Zilic, K. 2017,
The EBEX Balloon-Borne Experiment - Gondola, Attitude Control, and Control Software, ApJ (in press).

\bibitem[Then(1965)]{Then1965}
Then, J.W. 1965,
Bifilar Pendulum - An Experimental Study for the Advanced Laboratory,
Am. J. Phys., 33, 545-547.

\bibitem[Treilhou et al.(2000)]{TREILHOU20001423}
Treilhou, J.P., Coutelier, J., Thocaven, J.J. \& Jacquey, C. 2000,
Payload motions detected by balloon-borne fluxgate-type magnetometers,
Advances in Space Research, 26, 9, 1423-1426.

\bibitem[Yajima et al.(2009)]{Yajima2009}
Yajima, N., Imamura, T., Izutsu, N. \& Abe, T. 2009,
Scientific Ballooning: Technology and Applications of Exploration Balloons Floating in the Stratosphere and  the Atmospheres of Other Planets,
Engineering Fundamentals of Balloons, 15-75.

\end{thebibliography}
\end{document}